\documentclass[twocolumn,prb,showpacs,preprintnumbers,amsmath,amssymb]{revtex4}

\usepackage{graphicx}
\usepackage{dcolumn}
\usepackage{bm}
\usepackage{color}
\newcommand{\bk}{\textbf{k}} \newcommand{\bq}{\textbf{q}}
\newcommand{\bp}{\textbf{p}} 
\newcommand{\br}{\textbf{r}}

\begin{document}

\title{Sign inversion in the superconducting order parameter of LiFeAs inferred 
from Bogoliubov quasiparticle interference}

\author{Shun Chi$^{1,2,*}$, S. Johnston$^{1,2,4,*}$, G. Levy$^{1,2}$, S. Grothe$^{1,2}$, 
R. Szedlak$^{1}$, B. Ludbrook$^{1,2}$, Ruixing Liang$^{1,2}$, P. Dosanjh$^{1,2}$, 
S. A. Burke$^{1,2,3}$, A. Damascelli$^{1,2}$, D. A. Bonn$^{1,2}$, W. N. Hardy$^{1,2}$, 
and Y. Pennec$^{1,2}$
} 

\affiliation{$^1$Department of Physics and Astronomy, University of British Columbia, Vancouver BC, Canada V6T 1Z1} 
\affiliation{$^2$Quantum Matter Institute, University of British Columbia, Vancouver BC, Canada V6T 1Z4} 
\affiliation{$^3$Department of Chemistry, University of British Columbia, Vancouver BC, Canada V6T 1Z1}
\affiliation{$^4$Department of Physics and Astronomy, University of Tennessee, Knoxille, Tennessee 37996-1200, USA}

\date{\today}

\begin{abstract}
Quasiparticle interference, (QPI) by means of scanning tunneling
microscopy/spectroscopy (STM/STS), angle resolved photoemission spectroscopy
(ARPES), and multi-orbital tight binding calculations is used to investigate
the band structure and superconducting order parameter of LiFeAs. Using this
combination of techniques we identify intra- and interband scattering vectors between the
hole ($h$) and electron ($e$) bands in the QPI maps. Discrepancies in the band
dispersions inferred from previous ARPES and STM/STS are reconciled by
recognizing a difference in the $k_z$ sensitivity for the two probes. The
observation of both $h$-$h$ and $e$-$h$ scattering is exploited using 
phase-sensitive scattering
selection rules for Bogoliubov quasiparticles. From this we 
demonstrate an $s_\pm$ gap structure,
where a sign change occurs in the superconducting order parameter between the $e$ 
and $h$ bands. 
\end{abstract}


\maketitle

\section{Introduction}

A recurring theme in the study of unconventional superconductors is the pairing
of electrons via repulsive interactions,\cite{ScalapinoRMP,Mazin2008,Kondo,HirschfeldReview}   
rather than the attractive 
interaction mediated by phonons that occurs in conventional superconductors. Cooper pairing
driven by a repulsive interaction， such as exchange of antiferromagnetic spin
fluctuations， carries a distinguishing feature that the superconducting gap 
$\Delta(\bk)$ 
has changes in sign for different values of momentum $\bk$ in the Brillouin zone.
This means that identifying the gap symmetry and structure in the
high-temperature (high-T$_c$) iron pnictide superconductors is an essential step towards
understanding the origin of superconductivity in these 
systems,\cite{ScalapinoRMP,Mazin2008,Kondo,HirschfeldReview} as well 
as understanding their relation to the high-T$_c$ cuprates.\cite{ScalapinoRMP}

\begin{figure}
 \includegraphics[width=\columnwidth]{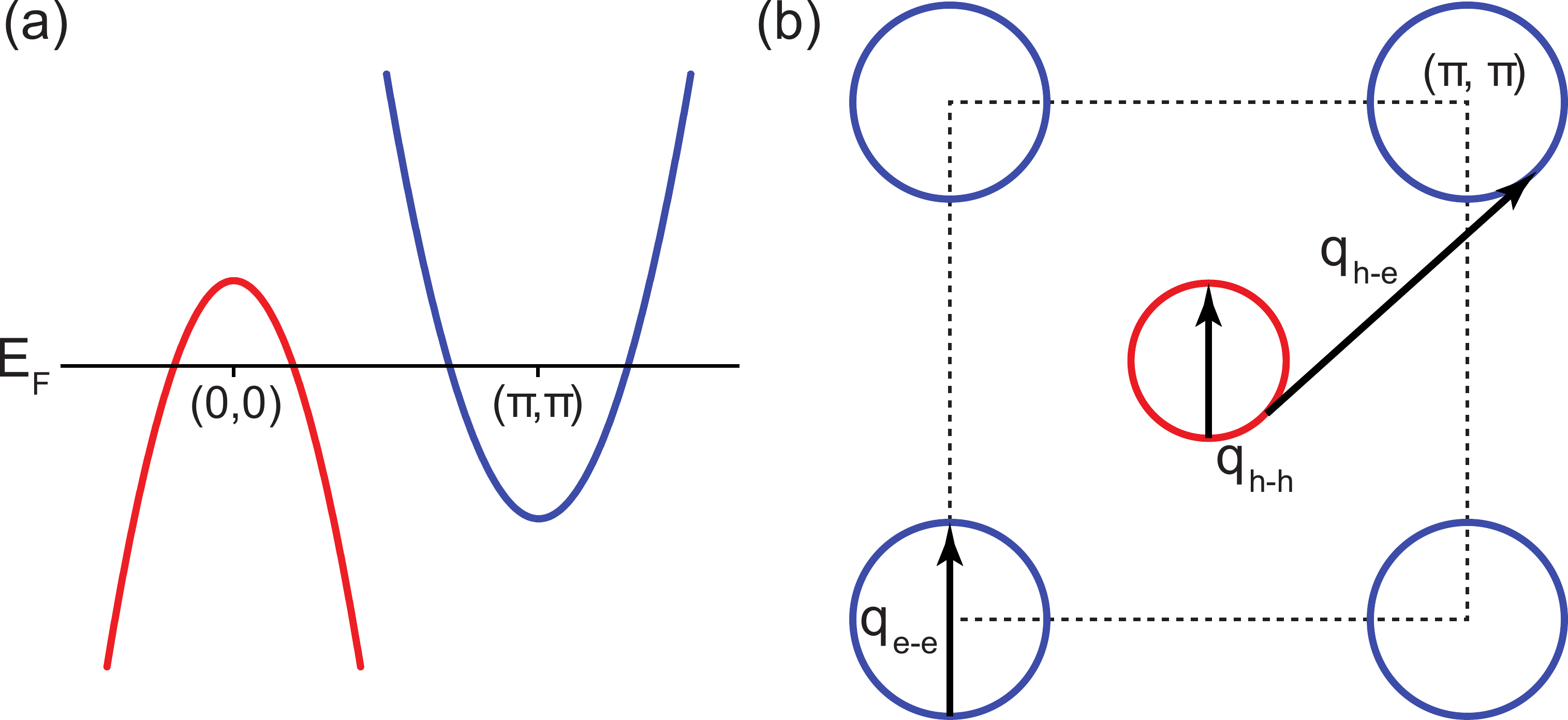}
 \caption{\label{Fig:1}
 (a) A simplified two-band model for the pnictides with a hole-like band
 centered at $\bk = (0, 0)$ and an electron-like band centered at $\bk = (\pi/a, \pi/a)$.
 (b) The Fermi surfaces of the bands in (a). The vectors $\bq_{h-h}$ and $\bq_{e-e}$ show
 intraband scattering within the hole and electron pockets, respectively, while
 $\bq_{h-e}$ shows interband scattering between the two. In the $s_\pm$ scenario, the sign
 of $\Delta(\bk)$ switches sign between the initial and final states of the $\bq_{h-e}$ 
 scattering process. 
}
\end{figure}

The electronic structure upon which superconductivity is built in the pnictides
consists of hole ($h$) bands centered at $\bk = (0,0)$ and electron ($e$) bands
centered at $\bk = (\pm\pi/a, \pm\pi/a)$ (Fig. \ref{Fig:1}a). In many 
pnictide systems the $h$ and $e$ bands are strongly nested leading to magnetic instabilities 
and superconductivity upon doping.\cite{Mazin2008} 
This nested multiband structure opens up the 
possibility that a sign change in momentum space could take the form of an $s_\pm$
gap structure, with $\Delta(\bk)$ having a different sign on the $e$ and $h$ 
bands. 
To date there has been considerable progress in measuring the gap
structures in iron pnictide and chalcogenide superconductors,
\cite{HirschfeldReview,HoffmanReview,Reid,Umezawa,DingEPL2008,HankePRL2012,BorisenkoSymm2012,
AllanScience2012,QureshiPRL2012,TaylorPRB2011,InosovNaturePhys2009,
HanaguriScience2010, ChristiansonNature2008, ChenNaturePhys2010,
CastellanPRL2011} including
many that can discern different gaps associated
with the multiple bands crossing Fermi level,
\cite{Umezawa,BorisenkoSymm2012,DingEPL2008,AllanScience2012,
QureshiPRL2012,TaylorPRB2011,Reid} and a few that are sensitive to whether or not
there is a sign change between the $e$ and $h$ bands.
\cite{CastellanPRL2011,InosovNaturePhys2009,HankePRL2012, HanaguriScience2010,
ChristiansonNature2008, ChenNaturePhys2010}
This $s_\pm$ symmetry is believed to be realized in the majority 
of pnictide superconductors. 

In this context, LiFeAs has a particularly important place amongst the pnictides. 
It is superconducting at its stoichiometric composition, enabling studies of the
superconducting state that are undisturbed by the disorder arising from
chemical substitutions.\cite{WangPRB2012,TappPRB2008,Pitcher2008,ChiPRL2012,GrothePRB2012}  
Furthermore, the system presents a natural
cleavage plane between two adjacent Li layers, which leads to stable, non-polar
cleaved surfaces with a carrier density similar to that of the 
bulk.\cite{LankauPRB2010} This
makes LiFeAs an ideal system for surface sensitive probes such as angle-resolved
photoemission spectroscopy (ARPES) 
\cite{BorisenkoPRL2010,KordyukPRB2011,BorisenkoSymm2012,Umezawa} and scanning tunneling
microscopy/spectroscopy (STM/STS), \cite{AllanScience2012,HankePRL2012, 
HessPRL2013,ChiPRL2012,GrothePRB2012,HanaguriPRB2012}  
and well suited for studying the underlying mechanism 
of superconductivity in the iron pnictides.

LiFeAs however differs from the 
other pnictides in some significant ways. 
The electronic structure of LiFeAs lacks the strong 
nesting conditions observed in other families, 
which is the likely reason for the absence of a magnetic 
phase.\cite{BorisenkoPRL2010,ZhegPreprint} Moreover, 
the underlying pairing symmetry in LiFeAs is under debate and it is unclear 
whether the nature of superconductivity in this material is the same as in the 
other pnictides. 

The lack of nesting between the $h$ and $e$ pockets weakens the traditional 
argument for $s_\pm$ pairing. This, coupled with the 
observation of multiple dispersion  
renormalizations, has led to proposals for an $s_{++}$ pairing symmetry 
\cite{BorisenkoSymm2012} driven by phonon assisted orbital fluctuations. 
\cite{TaylorPRB2011,KontaniPRL2010} 
Alternatively, proposals have been made for an exotic 
triplet pairing. For example, ARPES 
indicates the presence of a van Hove singularity at the top of the inner 
hole pockets,\cite{BorisenkoPRL2010} which can enhance ferromagnetic 
fluctuations and lead to a $p$-wave pairing symmetry. 
This is supported by a recent STM/STS study \cite{HankePRL2012} as well 
as theory based on the random phase approximation (RPA) and 
a two-dimensional (2D) three-band model.\cite{BrydonPRB2011} 
However, NMR and $\mu$SR measurements on high purity LiFeAs samples do not show 
any signature of triplet pairing.\cite{JeqlicNMR2009,LiNMR,MaNMR2010, BaekNMR2013, WrightmuSR2013}
In contrast, 
there are indications that an $s_\pm$ symmetry is
realized in LiFeAs despite the lack of strong nesting between the $h$ and $e$ 
bands. From a theoretical point of view, both an early functional renormalization group study 
(based on density functional theory (DFT) bandstructure calculations) \cite{PlattPRB2011} and a 
more recent RPA study (based on an ARPES-derived bandstructure) \cite{WangPreprint2013} 
find a leading $s_\pm$ superconducting instability. This scenario also has 
experimental support from a number of indirect probes.\cite{LiNMR,TaylorPRB2011,
QureshiPRL2012,DaiPRB}

The presence of a spin resonance mode observed by inelastic neutron scattering 
is perhaps the strongest indirect evidence in support of the $s_\pm$ gap symmetry in 
LiFeAs.\cite{ScalapinoRMP,EshrigReview}  
In general, a resonance peak occurs in the imaginary part of the spin susceptibility 
below T$_c$ at wave vector {\bf Q} connecting portions of the Fermi surface 
that have different signs for the superconducting gap. In LiFeAs, a broad 
resonance mode is observed at an incommensurate wavevector close 
to the antiferromagnetic wavevector connecting the $h$ and $e$ bands,\cite{TaylorPRB2011,
QureshiPRL2012,DaiPRB} consistent with an $s_\pm$ symmetry and similar to 
other pnictides.\cite{CastellanPRL2011,InosovNaturePhys2009,ChristiansonNature2008}
The observed energy scale $\Omega_{r}$ is consistent with 
a modulation in LiFeAs's LDOS,\cite{ChiPRL2012} 
indicating that this mode may be related to the pairing glue.\cite{ScalapinoRMP} 
There are however open questions regarding this interpretation. 
First, the observed spin resonance is rather broad  
in comparison to the sharp LDOS modulations.\cite{ChiPRL2012} Second, 
in the case of the high-T$_c$ cuprates it was shown that a
LDOS modulation in the form of a dip-hump feature is indicative of a 
pair breaking mode within the Eliashberg formalism.\cite{JohnstonPRB2010} 
Third, no corresponding feature has been observed in the ARPES spectra as 
would be expected if the mode was coupling strongly to carriers. 
In light of these issues we conclude that while the existence 
of a resonance mode is indicative of a sign change in the order parameter, 
its role in establishing superconductivity is not fully understood and its presence can only be considered as circumstantial evidence for an $s_\pm$ pairing symmetry. 

Given these open issues regarding the pairing symmetry in LiFeAs, 
it is desirable to have a direct, phase-sensitive measurement of the 
superconducting gap.  Here, we combine
ARPES, STM/STS, and multi-orbital scattering theory to study quasiparticle  
interference (QPI) in LiFeAs. Using this coherent approach to determine the electronic structure, 
we identify the relevant scattering vectors for this system and show that 
the energy dependence of the QPI intensity behaves as 
expected for an $s_\pm$ superconductor with scattering from a non-magnetic 
potential impurity. In this way we provide direct, 
phase-sensitive proof for an $s_\pm$ symmetry of the superconducting gap.  

The organization of this paper is as follows. In the following section we outline 
our experimental methods and the details of our model for QPI in LiFeAs. 
In section \ref{Sec:Results} we present results. 
We begin 
by summarizing our STM/STS results in section 
\ref{Sec:Results_QPI}.
In section \ref{Sec:vectors} we identify  
the scattering vectors in the QPI maps through detailed calculations 
based on an ARPES-derived bandstructure model for LiFeAs. 
Then, in section \ref{Sec:pairing} we examine the intensity variations of the QPI 
vectors and exploit a set of ``selection rules" to conclude an underlying $s_\pm$ pairing symmetry in this system. 
We then end in section \ref{Sec:Summary} with a summary and some concluding 
remarks.  We give the details of our 
data treatment for the QPI maps in appendix \ref{Sec:DataProcessing}. 

\section{Methods}\label{Sec:Methods}
\subsection{Materials and Experimental Details}\label{Sec:Exp}
Single crystals of LiFeAs (T$_c$ = 17.2 K) were grown by a self-flux technique 
\cite{ChiPRL2012,GrothePRB2012} first reported by
Morozov {\it et al.} (Ref. \onlinecite{Morozov}). For the STM/STS measurements, 
a LiFeAs single crystal sample was cleaved in-situ at cryogenic temperature
below 20 K and inserted into a beetle-type STM head operating
under ultrahigh vacuum (UHV) with pressure $P < 1\times10^{-9}$ Torr and at a base temperature of 4.2 K. The data was acquired at the STM base temperature 4.2 K which is the limiting factor for energy resolution. 
Electrochemically etched tungsten tips were used, which were Ar$^+$ sputtered,
and thermally annealed under UHV prior to measurement. 
The QPI data were obtained by numerical differentiation of the $I$-$V$ spectrum acquired at each pixel.

The ARPES measurements were performed with a SPECS Phoibos 150 analyzer and
21.218 eV linearly polarized photons from a monochromatized UVS300 lamp. The LiFeAs
single crystals were cleaved in-situ at a temperature of 6 K in an UHV 
environment with a base pressure of $P = 5\times10^{-11}$ Torr. The
full width at half maximum energy and angular resolutions were measured to
be 22 meV and 0.025$^{\circ}$, respectively. This corresponds to a momentum
resolution of 0.001$\pi/$a. 

\subsection{Multiorbital tightbinding model}\label{Sec:TB}
To model the electronic dispersion of LiFeAs we modified the ten orbital 
tight-binding model of Ref. \onlinecite{EschrigPRB2009}, 
formulated in two-Fe unit cell. 
In the normal state, the tight-binding Hamiltonian is given by 
\begin{center}
\begin{equation}
H_{ns}(\bk)=\sum_{\bk, \sigma}{\psi}_{\bk,\sigma}^{\dag}\
\hat{h}\left(\bk\right)\ {\psi}_{\bk,\sigma},
\label{eq:S2}
\end{equation}
\end{center}
where ${\psi{}}_{\bk,\sigma{}}^{\dag{}}=[c_{\bk,1,\sigma{}}^{\dag{}},\cdots,c_{\bk,10,\sigma{}}^{\dag{}}]$
is a row vector  of creation operators for the ten Fe orbitals. Here, we follow the 
notation of Ref. \onlinecite{EschrigPRB2009} and the matrix representation of 
the tight-binding Hamiltonian $\hat{h}(\bk)$ is given therein.

In order to obtain better agreement with the ARPES bandstructure at
$k_{z} = 0$, a handful of the hopping parameters were adjusted. (A comparison 
with the ARPES dispersion is shown in Fig. \ref{Fig:3}, and will be discussed 
in greater detail below.) 
Specifically  (in the notation of Ref. \onlinecite{EschrigPRB2009}), we set 
$\epsilon_{1}$ = -0.235, $\epsilon_{3}$ = $\epsilon_{4}$ = 0.23, $t_{18}^{10}$
= 0.211\textit{i}, $t_{27}^{10}$ = -0.258, $t_{33}^{11}$ = 0.267, 
$t_{34}^{11}$ = 0.0225, $t_{49}^{10}$ = 0.377, $t_{001}^{11}$ = 0.0714, 
and $t_{18}^{101}$ = $t_{19}^{101}$ = 0 (in units of eV). 
The remaining parameters remain unchanged from those
specified in the original model. Finally, the resulting bands were renormalized
by a factor of 2.17, which is typical for the iron pnictides.\cite{Mass} 

\subsection{Theory of Multiband Quasiparticle Interference}\label{Sec:QPI}
The QPI patterns are calculated using the usual T-matrix
formalism for a single impurity, formulated for a multiorbital system.\cite{ZhangPRB2009}  
The single impurity approach is justified by the dilute concentration of impurities
observed in our sample.\cite{ChiPRL2012,GrothePRB2012} 

First, it is convenient to establish some notation by  
introducing the band representation for the tight-binding Hamiltonian. We define $\hat{\epsilon}(\bk) = \hat{U}(\bk) \hat{h}(\bk) \hat{U}(\bk)^\dagger$, where $\hat{\epsilon}(\bk)$ 
is understood to be a $10\times10$ diagonal matrix whose diagonal elements are the 
eigenvalues of $\hat{h}(\bk)$ and $U(\bk)$ is the orthogonal transform  
diagonalizing $\hat{h}(\bk)$, which is obtained numerically.  
We introduce superconductivity in band representation by 
assigning a momentum independent instantaneous intraband pairing potential 
$\Delta_i(\bk) = \Delta_i$ to each band.  
The BCS Hamiltonian is then  $H_{bcs} = \sum_\bk \tilde{\Psi}_{\bk}^\dagger \tilde{B}(\bk) \tilde{\Psi}_{\bk}$ 
where $\tilde{\Psi}_{\bk}^\dagger = [\tilde{c}^\dagger_{\bk,1,\uparrow}, \cdots, 
\tilde{c}^\dagger_{\bk,10,\uparrow}, \tilde{c}^{\phantom{\dagger}}_{-\bk,1,\downarrow},\cdots,
\tilde{c}^{\phantom{\dagger}}_{-\bk,10,\downarrow}]$ 
and 
\begin{equation}
\tilde{B}(\bk) = \left[ \begin{array}{cc}
\hat{\epsilon}(\bk) & \hat{\Delta}  \\
\hat{\Delta} & -\hat{\epsilon}(-\bk)  \end{array} \right]
\end{equation}
is a $20\times 20$ matrix. 
Here, operators decorated with a tilde $\tilde{A}$ denote operators in band representation and both 
$\hat{\epsilon}(\bk)$ and $\hat{\Delta}$ are $10\times 10$ diagonal matrices whose 
$i$-th diagonal element is the eigenenergy $\epsilon_i(\bk)$ and pairing potential $\Delta_i$ 
for band $i$, respectively.    
Since the impurity must be introduced at the orbital level it is convenient to 
return to orbital representation by reinserting the orthogonal transformation 
$\hat{U}(\bk)$ 
\begin{equation*}
\hat{B}(\bk) = \left[ \begin{array}{cc}
\hat{U}^\dagger(\bk)\hat{\epsilon}(\bk)\hat{U}(\bk) & \hat{U}^\dagger(\bk)\hat{\Delta} \hat{U}^*(-\bk)  \\
\hat{U}^T(-\bk)\hat{\Delta} \hat{U}(\bk) & -\hat{U}^T(-\bk)\hat{\epsilon}(-\bk)\hat{U}^*(-\bk)  
\end{array} \right]
\end{equation*}
where $T$ denotes the transpose, $*$ the complex conjugate, and 
$\dagger$ the hermitian conjugate.

In orbital representation, the Green's function for the clean system 
in the superconducting state is given by 
\begin{equation}
\hat{G}_{0}(\bk,\omega) = [(\omega+i\delta)\hat{I} - \hat{B}(\bk)]^{-1},
\end{equation}
where $\delta$ is a broadening factor and $\hat{I}$ is the $20\times 20$ identity matrix. 
The impurity induced Green's function is given by 
\begin{eqnarray}\label{Eq:G}\nonumber
    \hat{G}(\bk,\bp,\omega)&=&\hat{G}_{0}(\bk,\omega)\delta_{\bk,\bp} + 
    \hat{G}_{0}(\bk,\omega)\hat{T}(\bk,\bp,\omega)\hat{G}_{0}(\bp,\omega) \\ 
    &=& \hat{G}_0(\bk,\omega)\delta_{\bk,\bp} + \delta\hat{G}(\bk,\bp,\omega),
\end{eqnarray}
where $\hat{T}$ is the T-matrix obtained by solving the matrix equation 
\begin{equation}
\hat{T}_{\bk\bp}(\omega) = \hat{V}_{\bk\bp} + 
\frac{1}{N}\sum_{\bk^\prime} 
\hat{V}_{\bk\bk^\prime}\hat{G}_{0}(\bk^\prime,\omega)\hat{T}_{\bk^\prime\bp}(\omega). 
\end{equation}

We consider the LDOS modulations induced by a single impurity  
that replaces one of the Fe atoms in the two-Fe unit cell.  For simplicity we assume that 
the potential scatterer affects all orbitals on the Fe site in the same way.  
The impurity Hamiltonian is given by 
\begin{equation}
H_{imp} = \sum_{i=1}^5 \sum_{\bk,\bp,\sigma} V_0 c^\dagger_{i,\bk,\sigma}c_{i,\bp,\sigma},
\end{equation}
where the sum over $i$ runs over the five orbitals on one of the Fe sites. 
Under these assumptions, the $T$-matrix is momentum independent and given by 
\begin{equation}
    \hat{T}(\omega) = [\hat{I} - \hat{V}\hat{g}(\omega)]^{-1}\hat{V},
\end{equation}
where $\hat{g}(\omega) = \frac{1}{N}\sum_\bk \hat{G}_{0}(\bk,\omega)$ and 
\begin{equation}\label{Eq:V0}
\hat{V} = V_0\left[ \begin{array}{cccc}
\hat{I} & \hat{0} & \hat{0} & \hat{0} \\
\hat{0} & \hat{0} & \hat{0} & \hat{0} \\
\hat{0} & \hat{0} &-\hat{I} & \hat{0} \\
\hat{0} & \hat{0} & \hat{0} & \hat{0} 
\end{array} \right].
\end{equation}
Here, each element  of the matrix in Eq. (\ref{Eq:V0}) represents a $5\times 5$ matrix.
The Fourier transform of the impurity induced LDOS modulations $\delta\rho(\bq,\omega)$ is then 
given by the trace over the imaginary part of $\delta\hat{G}(\bk,\bp,\omega)$ 
\begin{equation}
\delta\rho(\bq,\omega) =\frac{i}{N}\sum_{\bk} 
\sum_{i=1}^{10} \left[\delta\hat{G}_{ii}(\bk,\bk+\bq,\omega) - 
\delta\hat{G}^*_{ii}(\bk+\bq,\bk,\omega) \right].
\end{equation}
For our calculations we took $V_0 = 50$ meV, however our conclusions are not 
sensitive to this value.  Furthermore, we assumed superconducting gap 
values of $\Delta_{h_1} = \Delta_{h_2} = 7$ meV, $\Delta_{h_3} = 3$ meV, and 
$\Delta_{e_{1,2}} = -4$ meV.\cite{BorisenkoSymm2012,Umezawa} Note that since we restrict our simulations to energies 
above the gap edges, the precise choice in $\Delta_i$ values is not critical to our 
identification of the QPI wavevectors.

\subsection{Phase sensitivity}
STM/STS provides access to the phase of the superconducting gap by imaging QPI
of Bogoliubov quasiparticles, which are a superposition of $e$ and $h$ excitations.  
The QPI patterns are imaged in real space by measuring the differential conductance
$dI/dV$ between the tip and sample as a function of position ${\bf r}$ and 
energy $E$ (Fig. \ref{Fig:2}a). 
A Fourier transform of this image produces a $\bq$-space QPI intensity map
(Fig. \ref{Fig:2}c), where peaks occur at wave vectors connecting segments of the band
structure (Fig. \ref{Fig:1}b).\cite{HoffmanReview}  

The phase sensitivity arises from the coherence factors $u_i(\bk)$ and $v_i(\bk)$, 
which determine the degree of $e$ and $h$ admixture ($i$ is a band index). 
This is most easily understood by examining the scattering rate between initial 
and final states as determined by Fermi's golden rule.  
In the multiband superconductor the scattering rate for 
transitions between band $i$ and $f$ is proportional to \cite{Tinkham}
\begin{eqnarray} \label{Eq:FGR}
W_{i\rightarrow f}(\bk,\bk^\prime)&\propto & |u_i(\bk)u^{*}_f(\bk^\prime)\pm v_i(\bk)
           v^{*}_f(\bk^\prime)|^{2}\times \nonumber \\ 
          &&\quad |V(\bk^\prime-\bk)|^{2}N_i(\bk)N_f(\bk^\prime),
\end{eqnarray}
where $V(\bq)$ is the scattering potential at vector $\bq=\bk^\prime-\bk$ 
and $N_i(\bk)$ the partial density of states of band $i$. 
The negative and positive signs in Eq. (\ref{Eq:FGR}) correspond 
to scattering from a potential and a magnetic impurity, respectively. 
The phase of $\Delta_i(\bk)$ enters via  
the Bogoliubov coherence factors
\begin{eqnarray}
    v_i(\bk) &=& {\rm sign}(\Delta_{i}(\bk))\sqrt{\frac{1}{2}
    \left(1-\frac{\epsilon_i(\bk)}{E_i(\bk)}\right)}; \nonumber \\
    u_i(\bk) &=& \sqrt{1-|v_i(\bk)|^2}.
\end{eqnarray}
The term $|u_i(\bk)u^{*}_f(\bk^\prime)\pm v_i(\bk) v^{*}_f(\bk^\prime)|$ is close to $1$ for energies well outside the superconducting gaps, independent of the pairing phase. Close to or below the superconducting gap, where Bogoliubov quasiparticles play an important role, $v_i(\bk)$ 
and $u_i(\bk)$ become comparable in magnitude. Thus the term 
$|u_i(\bk)u^{*}_f(\bk^\prime)\pm v_i(\bk) v^{*}_f(\bk^\prime)|$ becomes a 
momentum-dependent prefactor differing from $1$ depending on the 
relative sign of 
$\Delta(\bk)$ and $\Delta(\bk^\prime)$.
This establishes a set of ``selection rules" that will enhance or suppress 
the scattering rate near the superconducting edge relative to 
the rate measured above it. Thus the QPI intensity is senstive to  
the nature of the impurity and the relative sign of $\Delta_i(\bk)$ and $\Delta_f(\bk^\prime)$.
These selection rules can also be more rigorously derived using 
the $T$-matrix formalism.\cite{SykoraPRB2011, DasJPCM2012}


\begin{table}
\begin{center}
\begin{tabular}{ | c | c | c |}
 \hline
 Scenario & $\bq$  & $\bq$  \\ 
 &  Suppressed Intensity &  Enhanced Intensity \\ \hline
 non-mag. imp., $s_{++}$ & $\bq_{h-h}$, $\bq_{e-e}$, $\bq_{h-e}$ & - \\ \hline
 mag. imp., $s_{++}$ & - & $\bq_{h-h}$, $\bq_{e-e}$, $\bq_{h-e}$ \\ \hline
 non-mag. imp., $s_{\pm}$ & $\bq_{h-h}$, $\bq_{e-e}$ & $\bq_{h-e}$ \\ \hline
 mag. imp., $s_{\pm}$ & $\bq_{h-e}$ & $\bq_{h-h}$, $\bq_{e-e}$ \\ \hline
\end{tabular}
\caption{\label{Tbl:1}  A summary of the QPI selection rules expected for a 
pnictide superconductor with $s_{++}$, $s_\pm$.  The QPI intensity of a scattering vector is either suppressed or enhanced inside the superconducting gap relative to the intensity outside the gap. The intensity variations stem from the energy dependence of the prefactor in Eq. \ref{Eq:FGR}. The four combinations of two pairing symmetries and two kinds of impurities result in 
four distinct sets of selection rules that can uniquely identify the 
pairing symmetry and the nature of the impurity.  
}

\end{center}
\end{table}

The selection rules for the pnictide band structure shown in Fig. \ref{Fig:1} 
are summarized in Table \ref{Tbl:1} for the cases of an $s_{++}$ and $s_{\pm}$ 
pairing symmetry. For instance, in the $s_\pm$ scenario with
non-magnetic impurities one expects the intensity of QPI vectors associated
with interband scattering between the hole and electron bands $\bq_{h-e}$ to be
enhanced while intraband scattering within the hole bands or the electron bands
$\bq_{h-h}$ and $\bq_{e-e}$, respectively, is suppressed when sweeping energies from above to inside the superconducting gap. 
Finally, we emphasize here that both the symmetry and nature of the impurity can be uniquely inferred from relative behaviour of subsets of the QPI intensities as indicated in Table \ref{Tbl:1}.

\section{Results}\label{Sec:Results}
\subsection{QPI Maps}\label{Sec:Results_QPI}
\begin{figure}
 \includegraphics[width=\columnwidth]{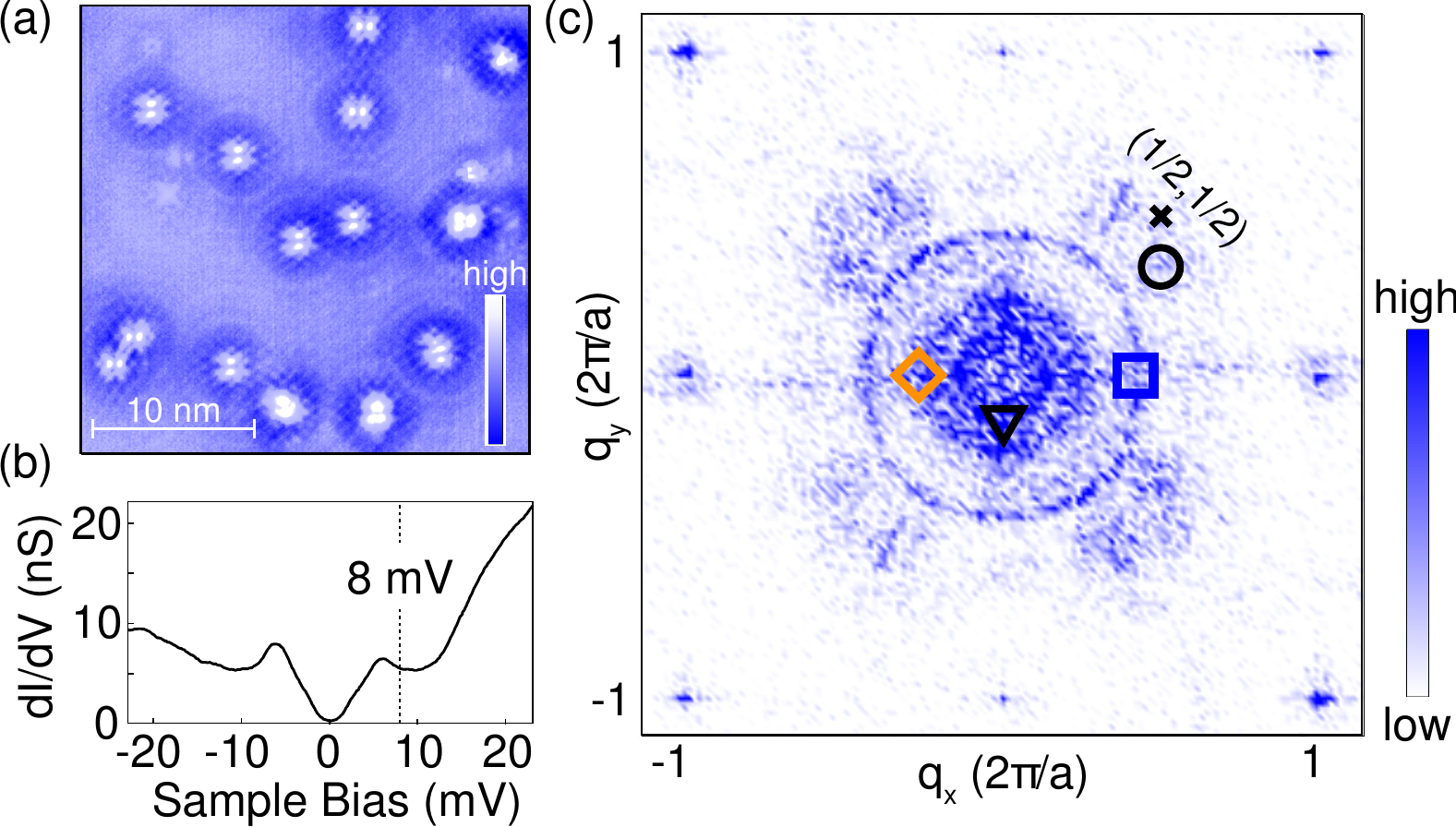}
 \caption{\label{Fig:2} (color online) 
  (a) A $26\times26$ nm$^2$ $dI/dV$ tunneling
 conductance map at $V_s = 8$ mV, which is outside the superconducting gaps.
 Fourteen distinct defects are observed, ten of which are Fe-D$_2$ defects. (b) A
 typical $dI/dV$ spectrum out of the original $400 \times 400$ pixel grid, selected from
 a pristine area. The rise in $dI/dV$ at $V_s > 12$ mV is a reproducible feature
 across every sample measured. (c) The QPI map obtained from a filtered 
 Fourier transform of the conductance map. (Filtering procedures are detailed in Appendix \ref{Sec:DataProcessing}.) Scattering among the hole
 bands appears as rings centered at $\bq = (0, 0)$ while interband scattering
 between the hole and electron bands appears as arcs centered around $\bq = (\pm\pi/a,
 \pm\pi/a)$. The symbols in (c) indicate the location of the QPI vectors whose
 dispersion is tracked in Fig. \ref{Fig:3}.
 }
\end{figure}
A summary of our STM/STS measurements ($T = 4.2$ K) is given in 
Fig. \ref{Fig:2}. 
Fig. \ref{Fig:2}a shows a $26\times26$ nm$^2$ tunneling conductance map 
of our sample taken at $V_s = 8$ mV. Fig. \ref{Fig:2}b 
shows a typical $dI/dV$ spectrum at a location far from any defect.
A clear $\Delta = 6$ meV superconducting gap is resolved along with a subtle shoulder
at $\sim 3$ meV. These values are consistent with the full double gap structure
found in the same sample at lower temperature ($T = 2$ K).\cite{GrothePRB2012,ChiPRL2012}  
A rapidly decaying diffraction pattern is present around each defect
(Fig. \ref{Fig:2}a), resulting from modulations of the local density of states (LDOS) due to quasiparticle scattering. The
corresponding QPI map (the two-dimensional power spectrum of Fig. \ref{Fig:2}a) 
is shown in Fig. \ref{Fig:2}c. Here we have applied a real-space and
momentum-space Gaussian mask method to remove a signal arising from the defect
centers that obscures the QPI intensity. The details of this procedure are given in 
Appendix \ref{Sec:DataProcessing}. No symmetrization has been applied to the data.  
Therefore, the symmetry of our QPI intensity map is certain to reflect the original 
symmetry of the underlying electronic structure.
\begin{figure*}[t]
 \includegraphics[width=\textwidth]{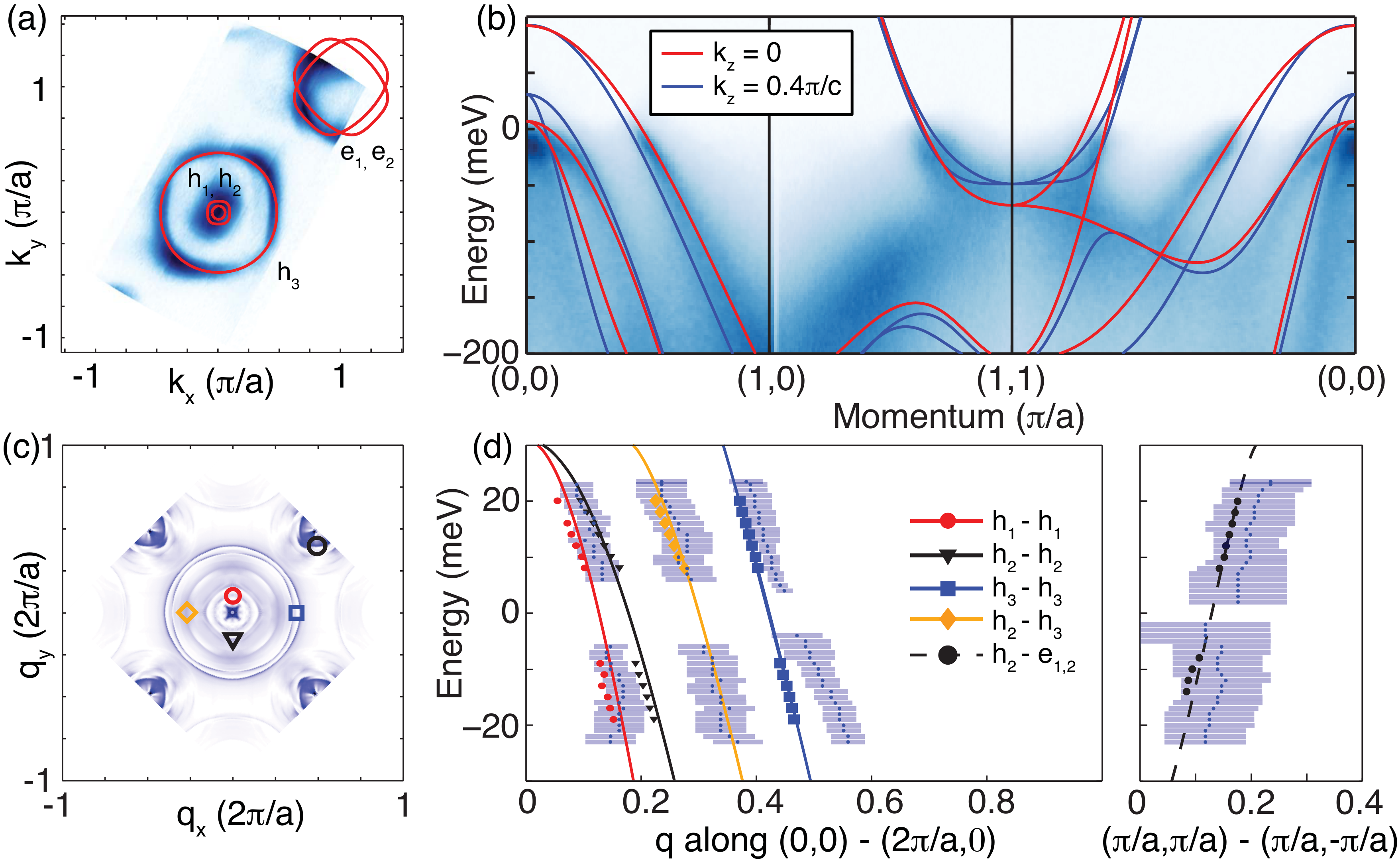}
 \caption{\label{Fig:3} (color online) 
 (a) The
 Fermi surfaces obtained from ARPES (shaded) and the model (solid lines). (b) The ARPES and model
 dispersions along the high-symmetry cuts of the first Brillouin zone. In (a)
 and (b) the ARPES spectra are shown for a photon energy that selects $k_z$ values
 near zero. The model dispersions are shown for $k_z = 0$ (red) and $k_z = 0.4\pi/c$
 (blue). (c) The calculated QPI at $V = 8$ mV assuming that electrons
 tunnel into a non-zero $k_z = 0.4\pi/c$. Features associated with intra- and
 interband transitions are indicated by the open symbols. (d) The experimental
 (blue points with error bars) and theoretical (solid symbols) dispersion of the
 QPI vectors indicated in (c). The error bars are determined approximately by the full width at half maximum of the QPI features plus one additional pixel uncertainty. The solid lines show the dispersion expected from the model dispersion.
}
\end{figure*}

The Bragg diffraction peaks of the As/Li $[(2\pi/a,0)]$ and Fe $[(2\pi/a,2\pi/a)]$ 
sublattices are clearly resolved at the outer edge of the QPI map. In addition
to the Bragg peaks, we find three features centered on $\bq = (0, 0)$; two small
inner rings and a larger outer ring, in agreement with previous 
studies.\cite{AllanScience2012,HankePRL2012}
We also observe a set of ``arc" features located midway along the $(0,0)-(\pm2\pi/a,
\pm2\pi/a)$ directions. Similar features were observed in Ref. 9, but these features were
at the edge of the data presented. These rings and arcs originate from multiple
inter- and intraband scattering processes, and due to the complexity of the
multiband electronic structure, their specific assignment has been
controversial.\cite{AllanScience2012,HessPRL2013} Allan {\it et al.}, 
(Ref. \onlinecite{AllanScience2012}) assigned the three rings to
intraband scattering within the three $h$ bands. However, the size of the bands
inferred from this interpretation is inconsistent with ARPES measurements. This was pointed
out by Hess {\it et al.} (Ref. \onlinecite{HessPRL2013}), who interpreted the inner and
outer rings as intraband scattering within two $h$ bands, and the middle ring as
interband scattering between the two. No assignment for the arc-like features
has been made to date. 

\subsection{Identification of the Scattering Vectors}\label{Sec:vectors}
To identify the underlying bands associated with each of these vectors, 
QPI maps were modeled using the $T$-matrix formalism 
outlined in section \ref{Sec:QPI}.   
In order to accurately identify each of the vectors observed 
it is important to anchor the model electronic structure to  
the empirical band structure observed by ARPES, as shown in Figs. 
\ref{Fig:3}a and \ref{Fig:3}b. With a photon energy of 21.2 eV, 
and based on an inner potential $V_0$ = 15.4 eV,\cite{HajiriPRB2012}
ARPES maps the electronic excitations for $\bk_{||}$ spanning the first Brillouin zone
at the average perpendicular momentum $k_z = 2.93\times 2\pi/c$, where $c = 6.31$ \AA\ 
is the lattice parameter perpendicular to the $(100)$ 
surface.\cite{Pitcher2008} This selects a $\bk_{||}$ 
plane intersecting the three-dimensional dispersion close to the $\Gamma$ 
point [up to a reciprocal lattice vector ${\bf G} = (0,0,6\pi/c)$, 
or $k_z \sim 0$ in a higher Brillouin zone].\cite{Footnote}  
The Fermi surface along this
$\bk_{||}$ plane (Fig. \ref{Fig:3}a) is composed of two hole pockets centered at 
$\Gamma$ (denoted $h_2$ 
and $h_3$) and two electron pockets centered at each of the $M$ points (denoted $e_1$ 
and $e_2$). A momentum distribution curve analysis of the ARPES spectra indicates 
the presence of a third inner hole pocket $h_1$, with 
the tops of the $h_1$ and $h_2$ bands located within a $\pm6$ meV window of $E_F$.
To model this electronic structure, we adopted the modified two-Fe ten-orbital
tight-binding model introduced in section \ref{Sec:TB}. The band dispersion 
for this model is shown in Figs. \ref{Fig:3}a and \ref{Fig:3}b.  

The calculated QPI intensity map at $V = 8$ mV, based on this model of the 
band structure, is shown in Fig. \ref{Fig:3}c, 
where we have assumed that electrons tunnel into a non-zero $k_z = 0.4\pi/c$ 
cut of the three-dimensional band structure.\cite{EschrigPRB2009,HajiriPRB2012}
(We will return to this point shortly.) A number of QPI vectors are 
present in the calculation, and are
highlighted by the open symbols. The calculation identifies the two innermost
rings (red $\bigcirc$ and black $\bigtriangledown$) 
and the outermost large ring (blue $\square$) with intraband scattering between three hole
bands, $h_1$-$h_1$, $h_2$-$h_2$, and $h_3$-$h_3$, respectively. The third ring 
from the center (orange $\Diamond$) is due to  
interband scattering between the inner and outer hole bands $h_1$-$h_3$ and 
$h_2$-$h_3$. Our model also identifies the arc-like features (black $\bigcirc$) 
centered on ($\pm\pi/a, \pm\pi/a$) with
scattering between the $h_2$ and $e_{1,2}$ bands. Scattering between the $h_3$ and 
$e_{1,2}$ bands is suppressed due to a mismatch of orbital character 
in these two bands.
\begin{figure*} 
 \includegraphics[width=0.90\textwidth]{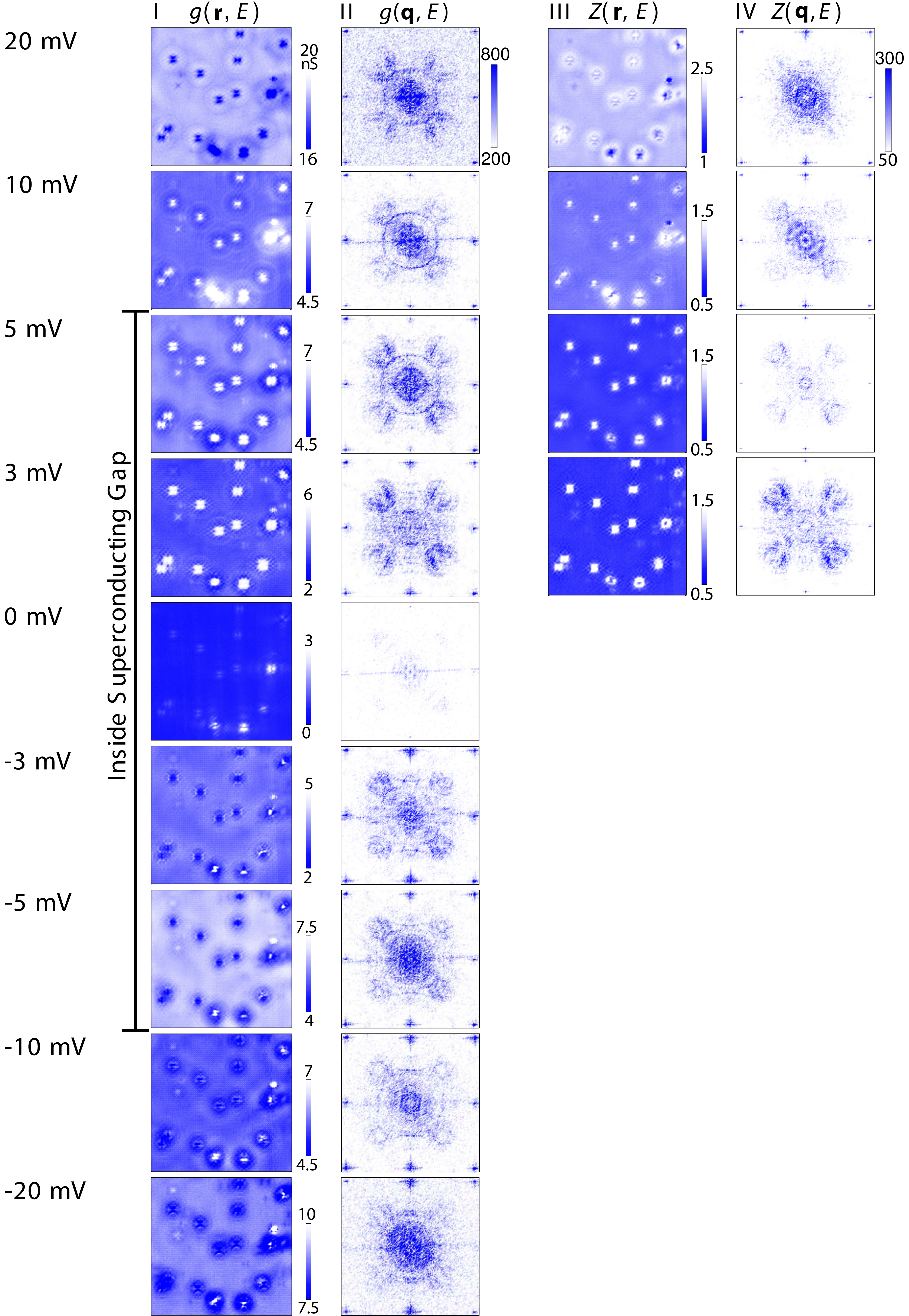}
 \caption{\label{Fig:4} 
 The energy dependence of the conductance maps $g(\br, eV)$ (column I) 
 and their corresponding Fourier transforms $g(\bq, eV)$  (column II). In computing
 $g(\textbf{q}, eV)$ at each energy we use the same parameters for
 Gaussian mask and Gaussian suppression (see appendix \ref{Sec:DataProcessing}). In addition, 
 QPI images in the second column are all plotted on the same color scale so that 
 the relative intensities can be directly compared. The corresponding 
 real space Z-maps defined as $Z(\br,eV) = g(\br, eV)/g(\br,-eV)$ are shown in column III, 
 while the Fourier transform $Z(\bq,eV)$ are shown in column IV in the same color scale. 
} 
\end{figure*}

Comparing to the experiment we associate the smallest to largest of the three
QPI rings in Fig. \ref{Fig:2}c with $h_2$-$h_2$, $h_2$-$h_3$, and $h_3$-$h_3$ scattering, 
respectively, and the arcs
with $h_2$-$e_{1,2}$ scattering.  
It should be noted that the consistency between the simulation with only an intraorbital 
scattering potential (see Fig.\ref{Fig:3}c) and the experimental QPI image (see Fig.\ref{Fig:2}c) 
implies that quasiparticle scattering primarily occurs between states with the 
same orbital character. These assignments are qualitatively consistent with 
Ref. \onlinecite{HessPRL2013}, however, the dispersion of the QPI vectors   
quantitatively disagrees with the ARPES band dispersion near the 
$\Gamma$-point (again, measured here a point related to 
$\Gamma$ by a reciprocal lattice vector). Notably, our ARPES measurements indicate 
that the top of $h_2$ is no more than 6 meV above $E_F$ at $\Gamma$; 
above this energy the $h_2$-$h_2$ 
and $h_2$-$h_3$ features should vanish due to phase space constraints if STM is 
probing the bandstructure in the $k_z = 0$ plane. 
This is
inconsistent with the observed QPI dispersions, shown in Fig. \ref{Fig:3}d (data points
with error bars) where all of the rings disperse to energies $> 20$ meV. In ARPES
experiments the value of $k_z$ can be controlled via varying the incident photon
energy and our photon energy corresponds to $k_z\sim 0$. However, in the case of
STS/STM, it is less clear which values of $k_z$ are probed in a bulk 3D system.
Empirically we have found that $k_z = 0.4\pi/c$ provides good agreement between our
model and the data (Fig. \ref{Fig:3}d). The solid symbols plot the dispersion of the
calculated QPI features. The agreement between the model and the experiment is
good and a non-zero value of $k_z$ reconciles differences in band structure
inferred from ARPES and STM/STS measurements.\cite{HessPRL2013,AllanScience2012,
BorisenkoSymm2012} We note that the agreement may be further improved by integrating signal over a range of $k_z$ values. However, this would required an explicit calculation of the tunneling matrix element and is left for future work. 

The fact that the inner hole pocket(s) disperse well above 20 meV at finite $k_z$ implies that a weak nesting condition exists between small inner hole pockets and comparatively large electron pockets at the Fermi level. This is consistent with a weak and incommensurate spin resonance mode revealed by inelastic neutron scattering (INS)  at a wavevector linking the $h$ and $e$ pockets.\cite{TaylorPRB2011,QureshiPRL2012}  
Furthermore, we observe a distinct scattering process
between the hole and electron Fermi surface sheets $h_2$-$e_{1,2}$ which, unlike the similar feature in Fe(Se,Te)\cite{HanaguriScience2010}, is well separated from the commensurate ($\pi, \pi$) point. This allows us to unambiguously disentangle QPI of $h$-$e$ scattering from Bragg peaks of possible charge or magnetic ordering.\cite{Mazin_Comment}

\subsection{Variation of the QPI Intensity and s$_\pm$ Pairing}\label{Sec:pairing}
Now that the QPI vectors have been identified, we turn to identifying 
the symmetry of the order parameter. This is accomplished by an examination of 
QPI of Bogoliubov quasiparticles near the superconducting gap, where the selection rules discussed 
in section 2 become dominant. They are reflected in the intensity variations 
in the QPI maps.  The energy dependence of the conductance maps and QPI intensities 
are shown in Fig. \ref{Fig:4}. The first and second columns 
of Fig. \ref{Fig:4} show the real and momentum space QPI maps. 
For completeness, the third column shows the so-called 
$Z$-maps in real space, defined as the ratio of the conductance maps 
at positive and negative biases $Z({\bf r},E) = g({\bf r},E)/g({\bf r},-E)$.
\cite{HanaguriNaturePhys2007}  
The fourth column is the corresponding Fourier transform $Z(\bq,E)$. 

\begin{figure}
 \includegraphics[width=0.7\columnwidth]{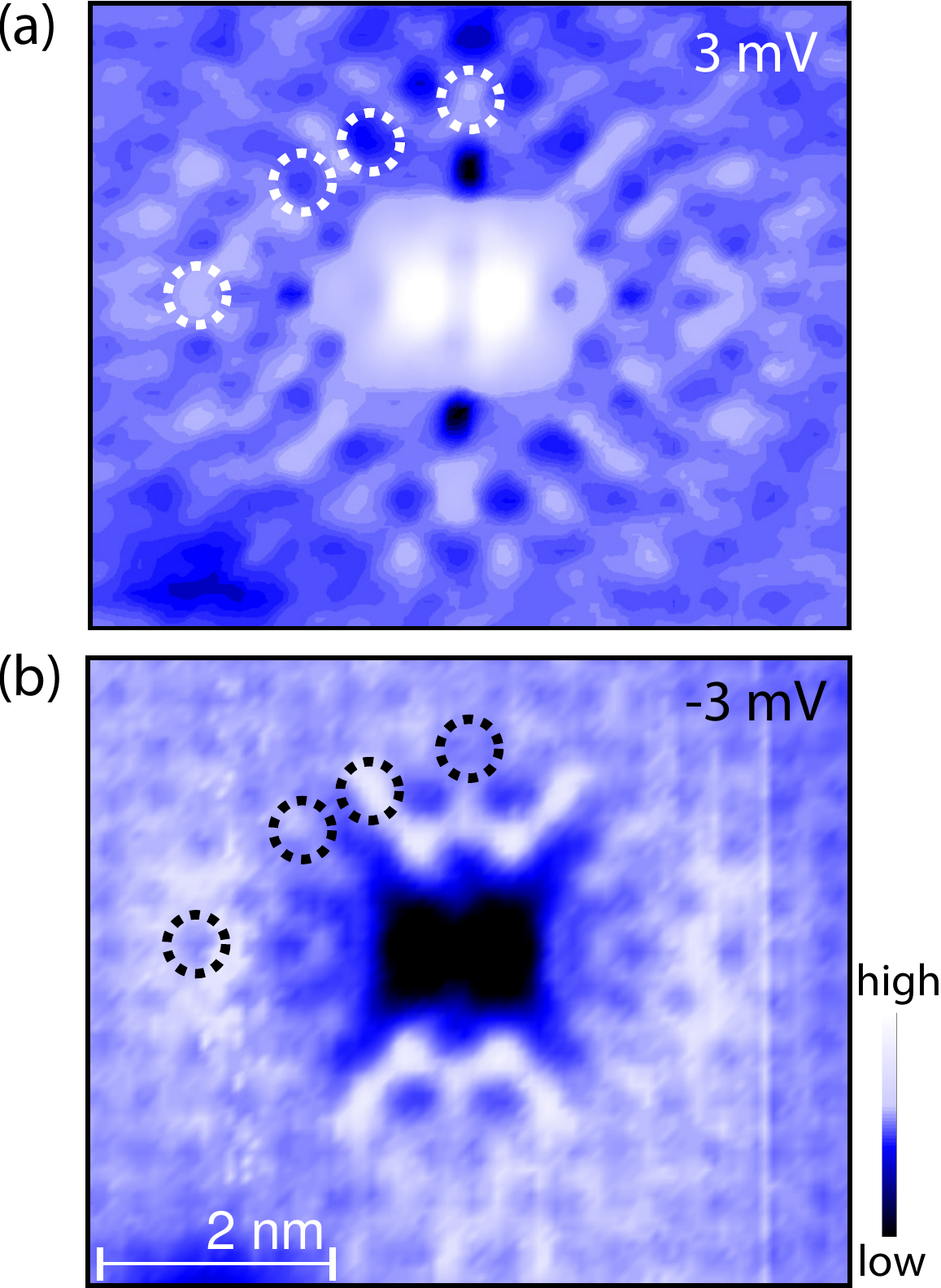}
 \caption{\label{Fig:7} (color online) Tunneling conductance maps
 $dI/dV$ in the vicinity of an Fe-D$_2$ defect at (a)
 $V_s = +3$ mV and (b) $V_s =-3$ mV. The anti-phase relationship of 
 the local density of states modulations is highlighted by the 
 locations with high intensities inside the white circles of (a) 
 and the corresponding low intensities inside black circles of (b), 
 and vice versa.}
\end{figure}

For biases well above the superconducting gap, 
the intensities of different scattering vectors in the QPI 
maps are relatively energy-independent; for example, compare the 
QPI maps at $V = 10$ and 20 mV. In contrast, 
as the bias voltage sweeps from above the gap (Fig. \ref{Fig:4} $g(\bq, eV)$ at 10 meV ) to 
inside the gap (Fig. \ref{Fig:4} $g(\bq, eV)$ at 3 mV), the intensity of the intra- and 
interband $h$ scattering is strongly
suppressed while the interband $h$-$e$ scattering is significantly enhanced. 

QPI of Bogoliubov quasiparticles 
distinguishes itself from normal state QPI by an anti-phase relation of LDOS
modulations at positive and negative energies. 
\cite{HanaguriNaturePhys2007,FujitaPRB2008} We illustrate this in
Fig. \ref{Fig:7}a and \ref{Fig:7}b, which show LDOS modulations near an Fe-D$_2$ 
defect at $\pm3$ mV.\cite{GrothePRB2012} 
The anti-phase relation is apparent in the contrast inversion highlighted inside the dashed circles in Fig. \ref{Fig:7}. This confirms the dominance of 
Bogoliubov QPI inside the superconducting gap.  Z-maps emphasize the anti-phase component of  
Bogoliubov QPI while suppressing the in-phase component of normal state QPI. \cite{HanaguriNaturePhys2007,FujitaPRB2008} 
$Z(\br, eV)$ at 3 mV (see Fig. \ref{Fig:4}) shows 
strong short-wavelength real-space modulations (column III) near each impurity. 
The Fourier transform $Z(\bq,E = 3\ \mathrm{meV})$ reveals strong 
intensity arcs near ($\pm\pi/a,\pm\pi/a$), corresponding to the previously identified 
$e$-$h$ scattering vectors. The intensity of these arcs diminishes as the bias voltage 
sweeps from inside the large gap to above it. Therefore the intensity variations 
observed in this energy range are indeed due to the selection rules imposed by the 
symmetry of the order parameter.

\begin{figure}
 \includegraphics[width=0.7\columnwidth]{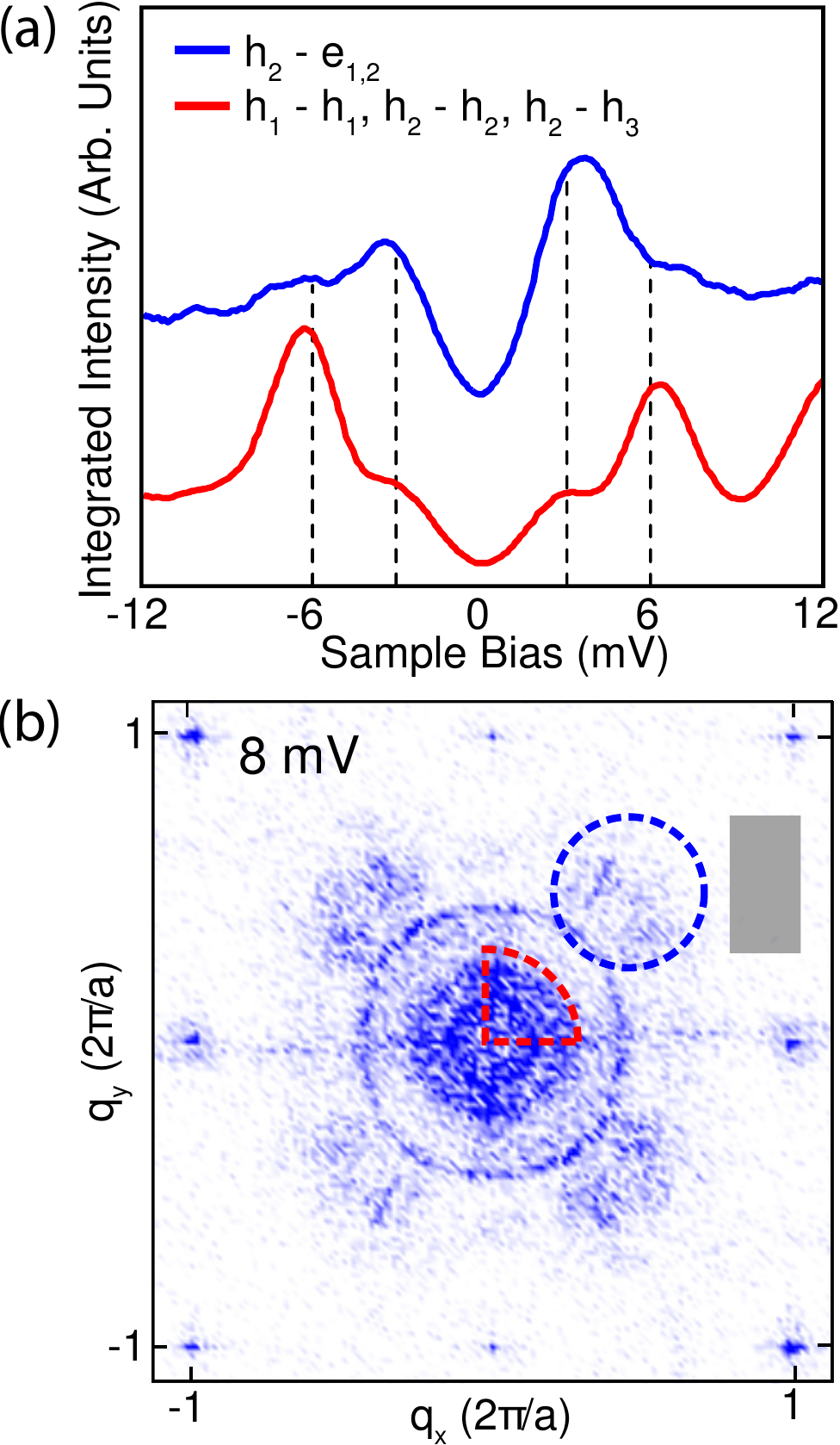}
 \caption{\label{Fig:5} (color online) (a)  
 The integrated intensity of the QPI signal for the
 intraband $h$-$h$ (red) and interband $h$-$e$ (blue) scattering vectors.
 The curves were then normalized to the value at 12 meV 
 and the interband intensity has been offset for clarity. The dashed lines indicate
 the values of the superconducting gaps. (b)
The red sector and blue circle are the integration windows for intraband $h$-$h$ and interband $h$-$e$ scattering intensities in panel (a), respectively. A noise background signal is integrated in the grey rectangular area and subtracted. Here the windows are shown in one quarter for simplicity but the integration is performed over the equivalent areas in all four quadrants of the image.
}
\end{figure}

We further quantify these 
intensity variations shown in Fig. \ref{Fig:4} by examining the integrated 
weight of each QPI vector as
a function of energy, as shown in Fig. \ref{Fig:5}a. 
Here, the intraband $h$-$h$ and interband
$h$-$e$ scattering vectors are isolated by defining appropriate integration windows
shown in Fig. \ref{Fig:5}b. 
The integrated intensity follows the behavior reflected in
the $dI/dV$ maps, demonstrating the selection rules for the entire data set. 

Comparing to the selection rules in Table \ref{Tbl:1}, we find that the data is most 
consistent with an $s_\pm$ scenario with $\Delta(\bk)$ changing sign 
between the $h$ and $e$ bands; below the superconducting gap $h$-$h$ scattering
intensities are suppressed while $e$-$h$ scattering intensities are enhanced for
non-magnetic impurities \cite{DasJPCM2012,ZhangPRB2009}. Magnetic impurities in the $s_\pm$ scenario have the opposite effect. The observed selection rules 
are also distinct from the $s_{++}$ scenario with either magnetic or non-magnetic 
impurities, see Table \ref{Tbl:1}. Furthermore, by considering that Knight shift decreases below $T_c$,\cite{JeqlicNMR2009,MaNMR2010, BaekNMR2013} we rule out the chiral p-wave state. 
Our results indicate the non-magnetic nature of the most common defect. This is consistent 
with the expectation that the Fe-D$_2$ defect is most likely a Li substitution 
on an Fe site, or an Fe vacancy,\cite{GrothePRB2012} both of which are expected to be 
non-magnetic. We therefore infer that the only candidate consistent
with our measurements is an $s_\pm$ symmetry. 

\section{Summary and Conclusions}\label{Sec:Summary}
We have examined QPI in LiFeAs using a combination of STM/STS, ARPES, and a tight binding model. By anchoring our tight binding description of LiFeAs to the ARPES-derived band dispersions we were able to unambiguously assign 
each of the scattering vectors in the QPI maps. In this framework, we have reconciled not only the discrepancies in the assignments of scattering vectors in prior QPI studies but also the disagreement on the sizes of inner hole pockets between ARPES and STM techniques by recognizing a non-trivial $k_z$ dependence in the 
tunneling process.  
With the assignment of the scattering vectors made, we then examined 
the detailed variations of the QPI intensity as a 
function of bias voltage. The variations in intensity 
near the superconducting gap are only consistent with 
an $s_\pm$ pairing symmetry where the change in sign occurs between 
the electron and hole pockets. Together with the observation of a spin fluctuation resonance by INS\cite{TaylorPRB2011,QureshiPRL2012}, this work presents a compelling evidence of unconventional $s_\pm$ pairing in LiFeAs driven by repulsive spin fluctuation interactions. 
This implies LiFeAs shares a common superconducting mechanism with the other members in the iron pnictide family.\cite{ScalapinoRMP,Mazin2008,Kondo,HirschfeldReview} Hence LiFeAs is a simple and clean model material for probing the common physics of iron pnictides. 

This work also demonstrates how Bogoliubov QPI from defect/impurity scattering provides a direct phase sensitive measurement of superconducting pairing symmetry. 
Bogoliubov QPI has been used to confirm the sign flip in the $d$-wave
Ca$_{2-x}$Na$_x$CuO$_2$Cl$_2$ cuprate superconductor\cite{HanaguriScience2009}
as well as Fe(Se,Te) iron-based superconductor\cite{HanaguriScience2010} under
high magnetic field, where vortices behaved as magnetic scattering centers.
However, this method is only suitable for materials with very short superconducting coherence lengths so that a vortex can be treated as a localized strong
magnetic scattering center. Here Bogoliubov QPI is measured (without the application of a magnetic field) by taking advantage
of point defects/impurities inside the material, which has been proposed only
theoretically.\cite{ZhangPRB2009, SykoraPRB2011, DasJPCM2012} This method can
be generalized to other superconductors, provided the nature (magnetic vs nonmagnetic) of the impurities are known.


{\it Acknowledgements:} S. C. and S. J. contributed equally to this work.  
The authors acknowledge I. Elfimov, C. Hess, P. J. Hirschfeld, A. Kemper, I. I. Mazin, G. A.
Sawatzky, E. van Heumen, and P. Wahl for discussions. This work was supported
by the Killam Foundation, Alfred P. Sloan Foundation, NSERC’s Steacie Memorial Fellowship (A.D.), the
Canada Research Chairs Program (A.D., S.B.), CFI, NSERC, and CIFAR Quantum
Materials. This work was made possible in part by the facilities of the Shared
Hierarchical Academic Research Computing Network and Compute/Calcul Canada.

\appendix
\section{Data Processing Methods}\label{Sec:DataProcessing}
\begin{figure*}[h!] \begin{center}
\includegraphics[width=\textwidth]{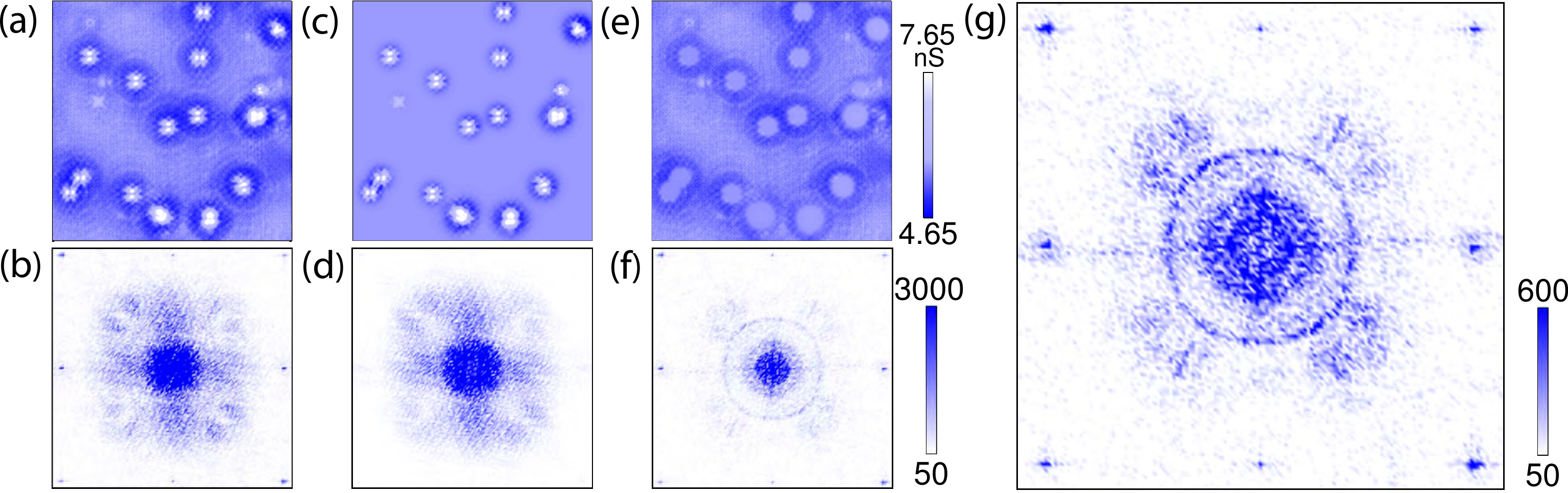}
\caption{\label{fig:S2} Examples of the processing techniques for the STM/STS
data. (a) The tunneling conductance map $g({\br}, 8\:\mathrm{meV})$ and
(b) its Fourier transform. (c) The portion of the 8 mV conductance map
removed from the defect centers and (d) its corresponding Fourier
transform. (e) Masked tunneling conductance map $g_{M}({\br}, 8\:
\mathrm{meV})$ and (f) its corresponding Fourier transform. (d) The
resulting QPI pattern.
Panels (a), (c) and (e) are in the same color scale as indicated next 
to panel (e); 
(b), (d) and (f) are in the same color scale as indicated next to (f). 
The major contribution to the raw signal in (a) comes from the defect
centers (b). 
Panel (g) shows the final QPI map after the additional application 
of the Gaussian suppression method of Ref. \onlinecite{AllanScience2012}.  
}
\end{center} \end{figure*}

In this appendix we outline our data processing method for the QPI maps shown 
in the main text.  
Figs. \ref{fig:S2}a and \ref{fig:S2}b show the tunneling conductance maps
$g(\br, V)$ at $V = 8$ mV and its direct Fourier transform
$g(\bq, V)$, respectively. Although there are obvious Friedel
oscillations around each defect, the Fourier transformed image does not show a
clear QPI pattern due to a dominant background signal centered at
\textbf{\textit{q}} = (0, 0). Here, we employ two methods to remove this
background and recover the underlying QPI patterns.

As shown in Fig. \ref{fig:S2}a, $g(\br, 8\:\mathrm{mV})$
exhibits strong conductance peaks at the defect centers that give rise to a
strong background signal in momentum space, overwhelming the QPI signal. In
general, the tunneling conductance map $g(\br, E
= eV) = dI/dV(\br, eV)$ is given by
\begin{center}
\begin{equation}
g\left(\br,\ eV\right)=\frac{eI_tN(\br,\ eV)}{\int_0^{eV_s}N\left(\br,\ E\right)\ dE}
\label{eq:S1}.
\end{equation}
\end{center}
where \textit{e} is a unit charge, $N(\br, eV)$ is the
LDOS at \textbf{r}, and \textit{E} = eV, and $V_s$ is the setting bias voltage for $I$-$V$ spectra.\cite{HanaguriNaturePhys2007} 
According to Eq. (\ref{eq:S1}) the variation of $g(\textbf{r}, E
= \mathrm{eV})$ is directly proportional to the variation of the LDOS if 
the normalization 
$\int_{0}^{\textit{eVs}}N(\textbf{r}, eV) \textit{dE}$ is
spatially homogeneous. This
condition however, does not hold at the center of the defects when the defects strongly
modify the local potential. This is because
$N(\textbf{r}, E)$
can be dramatically modified by local changes in the electronic structure and/or
the creation of bound states. The LDOS of LiFeAs is highly inhomogeneous near
E$_{F}$,\cite{ChiPRL2012} so the defect-induced changes in the local electronic structure
cause a significant variation in the integral of the LDOS over the energy 
range $[0,25 \:\mathrm{meV}]$.  
In addition, all common defects in LiFeAs generate bound states inside the
superconducting gaps.\cite{GrothePRB2012} Therefore, the behavior of
$g(\textbf{r}, E)$ at defect centers cannot be simply
interpreted as Friedel oscillations in $N(\textbf{r}, E)$ due
to the inhomogeneity of the normalization factor. 

A Gaussian masking method is used to eliminate the signal from the central
conductance peaks of these defects. For a defect located at
$\textbf{r}_{0}$, the masked conductance map $g_{M}(\textbf{r}, E)$ is given by
$g_{M}(\textbf{r}, E) = g(\textbf{r}, E) \times (1-M(\textbf{r}-\textbf{r}_{0},\sigma))$, 
where $M(\textbf{r}-\textbf{r}_{0},\sigma)$ is a truncated Gaussian
function with the maximum value = 0.99 and $\sigma$ is the standard deviation, 
taken to be approximately the half width of the defect center. This
Gaussian masking method suppresses the local conductance peaks associated with
the defect centers yet preserves the sign of $g(\textbf{r}, E)$ and produces a
smooth transition from the masked regions to the QPI nearby. 
We apply the Gaussian mask to each of the defects. Fig. \ref{fig:S2}c and \ref{fig:S2}d 
show the portion of the real space conductance map removed by the mask and its
corresponding Fourier transform, respectively. This demonstrates that the defect
centers primary contribute a large background signal centered at \textbf{q}
= (0, 0). Fig. \ref{fig:S2}e and \ref{fig:S2}f show the real space and momentum space
conductance maps after the application of the Gaussian mask, respectively. After
the removal of the defect center's background signal, significantly more
symmetric and regular patterns stand out in momentum space. 
A small but high intensity ring is present in the center and is
surrounded by a second less intense ring and a third larger ring. 
In addition, arc structures appear
in the direction $(0, 0) - (\pm 2\pi/a,\pm 2\pi/a)$.  In the analysis, no
symmetrization has been applied to the data. Thus any feature satisfying the
tetragonal crystal symmetry is real and originates from the underlying electronic
structure.  

As shown in Fig. \ref{fig:S2}f, the strong intensity around
$\textbf{q} = (0, 0)$ lowers the visibility of the QPI pattern at larger
\textbf{q}. We therefore further applied the Gaussian suppression method
of Allan \textit{et. al.} (Ref. \onlinecite{AllanScience2012}) to suppress the central peak:
$g(\textbf{q}, E) = g_{raw}(\textbf{q}, E)\times (1-0.95 \times G(\textbf{q}, \sigma))$, where 
$G(\textbf{q}, \sigma)$ is a Gaussian function with peak value = 1 and
$\sigma \sim 0.35 \pi/a$. We chose to retain 5\% of the signal at
$\textbf{q} = (0, 0)$ in order to not overly suppress the real QPI signal
near $\textbf{q} = (0, 0)$. Fig. \ref{fig:S2}g shows the final QPI map after
applying both the Gaussian mask in real space and Gaussian suppression in
momentum space. We emphasize that the same treatment with the same masking parameters was applied to all of the QPI maps.


\begin{thebibliography}{100}
\bibitem{Footnote}
By changing $\bk_{||}$ one also changes the corresponding $k_z$, such that the
measured momenta belong to a curved surface in $\bk$-space. However, with 21.2 eV
photons and $V_0 = 15.4$ eV, $k_z$ varies from $5.9\pi/c$ to $5.3\pi/c$ over 
the $\Gamma$ to $M$ range
in $k_{||}$, which allows us to approximate the measured surface simply as a 
$k_{||}$ plane.
\end{thebibliography}
\end{document}